\documentclass[12pt]{iopart}
\usepackage{epsfig}
\usepackage{graphicx}% Include figure files
\usepackage{dcolumn}% Align table columns on decimal point
\usepackage{bm}% bold math
\usepackage{color}% Emphasise changes
\begin{document}

\title{Gate-controlled charge transfer in Si:P double quantum dots}

\author{F E Hudson$^1$, A J Ferguson$^1$\footnote{Present address:
  Microelectronics Research Centre, Cavendish Laboratory, University of
  Cambridge, CB3 0HE, UK}, C C Escott$^1$, C
  Yang$^2$, D N Jamieson$^2$, R G Clark$^1$ and A S Dzurak$^1$}
\address{$^1$ Centre for Quantum Computer Technology,
Schools of Physics and Electrical Engineering, University of New
South Wales, NSW 2052, Australia}
\address{$^2$ Centre for Quantum Computer Technology, School of
Physics, University of Melbourne, VIC 3010, Australia}
\ead{f.hudson@unsw.edu.au}

\begin{abstract}We present low temperature charge
sensing measurements of nanoscale phosphorus-implanted double-dots
in silicon. The implanted phosphorus forms two 50~nm diameter
islands with source and drain leads, which are separated from each
other by undoped silicon tunnel barriers. Occupancy of the dots is
controlled by surface gates and monitored using an aluminium
single electron transistor which is capacitively coupled to the
dots. We observe a charge stability diagram consistent with the
designed many-electron double-dot system and this agrees well with
capacitance modelling of the structure. We discuss the
significance of these results to the realisation of smaller
devices which may be used as charge or spin qubits.
\end{abstract}

\maketitle

Few electron double-dot systems with controllable quantum
mechanical charge or spin states are of significant interest for
quantum information processing. To date, GaAs-based double quantum
dots have proved the most fruitful in achieving coherent control
\cite{Hayashi,PettaScience,KoppensNature}, however, there has been
recent progress in materials systems such as SiGe
\cite{Berer,Sakr,Simmons,Slinker}, carbon nanotubes
\cite{Biercuk,Sapmaz} and semiconductor nanowires
\cite{Lieber,Thelander,Zhong}. Silicon systems, for example those
based on phosphorus donors in silicon (Si:P), are particularly
appealing for single donor charge- and spin-based quantum
information processing \cite{Kane,Hill,Hollenberg}, promising long
spin coherence times \cite{tyryshkin2003} as well as compatibility
with existing silicon CMOS techniques. Consequently, there has
been a renewed focus on the realisation of Si-based single
\cite{Angus07} and double quantum dots
\cite{Gorman,Hofheinz,Fujiwara,Shin} in which Coulomb blockade and
double-dot charging diagrams have been reported.

In this paper, we study a silicon double-dot system where the dots
are defined by implantation of phosphorus donors in silicon to
form nanoscale, metallically-doped ($n^+$) islands electrically
isolated by regions of undoped Si. Our aim is to study Si:P
double-dots of reduced dimensions and explore the possibility of
realising double-dots in silicon that could operate as either charge
or spin qubits. We build on recent work in which we
studied  transport through a single-dot with $N~{\sim}$~600 donors
\cite{HudsonMNE}, charging of large double-dots containing
$N~{\sim}$~10,000 donors \cite{ChanJAP} and experiments that used
aluminium single electron transistors (SETs) to investigate the
transfer of electrons between two isolated islands of P donors in
silicon \cite{Buehler}.

The devices described here have two phosphorus-doped islands
separated by small gaps from ion implanted source ($S$) and drain
($D$) leads. These differ from the majority of other double-dot
systems which use surface gates to define the dots in an
underlying structure. The double-dot devices are designed to
contain either $N$~= 300 or 600 confined electrons in each dot,
with this number reduced slightly due to population of interface
traps. Occupancy of the dots is manipulated by surface control
gates but, due to the small dimensions of the device, the number
of gates is limited to two control gates to modify the
electrostatic potential of the dots. At this scale, it is
difficult to include tunnel barrier control gates between each dot
and the leads and so the device depends on geometrical design for
appropriately high tunnel barriers. The resulting tunnel coupling
is sufficiently weak to preclude direct transport measurements and
so charge transfers are monitored using an aluminium SET as a
charge detector.

The devices were fabricated on a high-resistivity silicon
substrate with $\rho >$ 5 k{$\Omega$}cm$^{-1}$ and a weak
background {\it n}-doping of less than $10^{12}$ ${\rm cm}^{-3}$.
(Figure 1(a) shows a completed device.) A 5~nm SiO$_{\rm 2}$ gate
oxide was thermally grown on the surface. This oxide was grown in
our laboratory and has a trap density of less than $n_{\rm trap} =
2 \times 10^{11}$ cm$^{-2}$, as determined from both MOSFET
threshold voltages and deep level transient spectroscopy (DLTS)
measurements \cite{McCamey_semi}.

The double-dots and source-drain leads were ion-implanted through
a polymer mask defined in 150~nm-thick PMMA resist by electron
beam lithography (EBL). The mask contained two 30~nm diameter
apertures to define the dots and openings for source-drain leads,
all separated by 100~nm gaps. For $N$ = 300 and 600 P atom dots,
phosphorus was implanted with areal doses of 4.3 and 8.5
${\times}~10^{13}$~cm$^{-2}$ at 14 keV energy, respectively,
resulting in a peak donor concentration at a depth of 20~nm below
the substrate surface. Based on modelling using an
industry-standard package \cite{SRIM}, we calculate peak densities
of the implanted regions to be
$n_{300}$~=~2.1~${\times}~10^{19}$~cm$^{-3}$ and
$n_{600}$~=~4.2~${\times}~10^{19}$~cm$^{-3}$, which are an order
of magnitude greater than the bulk metal-insulator transition for
Si:P ($n_{\rm MIT}$~=~3.45~${\times}~10^{18}$~cm$^{-3}$).

Following implantation, the PMMA resist mask was removed and the
implanted phosphorus regions imaged with a scanning electron
microscope (SEM). Figure 1(c) shows an SEM image of an implanted
device at this stage. In this image, the dark areas are due to the
presence of the implanted phosphorus donors and also the damage
caused by the ion implantation process. The individual dimensions
of each device can be recorded and we observe that dots implanted
through a 30~nm diameter aperture yield a damaged region of $\sim$
50~nm diameter. From this we infer the diameter of the implanted
dots to be $\sim$ 50~nm and the gaps between the dots and the
leads to be $\sim$ 80~nm. A rapid thermal anneal (RTA) performed
at 1000~$^{\circ}$C for 5~s repaired the damage caused by the
implantation process while minimising dopant diffusion
\cite{Oehrlein}. The contrast between the implanted region and the
undoped silicon upon SEM imaging decreases significantly after
this anneal (seen in Figure 1(a)), consistent with repair of the
implantation damage.

In two final EBL steps, electrostatic gates and the SET were
patterned on the silicon surface. Ti/Au control gates, $V_{\rm L}$
and $V_{\rm R}$, were fabricated using a single layer PMMA resist
mask and thermal metal evaporation. The SET was defined using a
bilayer polymer resist and double-angle evaporation. An in-situ
oxidation was performed between the two layers to form the oxide
tunnel barriers. Apertures, gates and SET were all aligned to one
another with better than 50~nm precision using Ti/Pt alignment
markers, chosen to withstand all processing steps, particularly
the high temperature anneal. The SET is positioned centrally
between the two dots which, despite being non-optimal for
inter-dot transitions, still allows us to detect transfers between
the dots due to a slight misalignment during fabrication.

The SET charge detection measurements were performed at the base
temperature of a dilution refrigerator (T ${\sim}$ 100~mK) using
standard low-frequency ($f$~$<$~100~Hz) lock-in techniques with an
ac excitation voltage $V_{\rm ac}$~=~20~$\mu$V. Whilst SET
sensitivity is highest in the superconducting state, the best data
was obtained when a magnetic field of B = 0.5~T was applied to
suppress superconductivity as fewer random
telegraph events were present in this state.

Devices with $N$ = 600 and 300 implanted P atoms per dot were
measured. A schematic cross section of the devices is shown in
Figure 1(b). The SET was biased at a point of high charge
sensitivity whilst the control gates, $V_{\rm L}$ and $V_{\rm R}$,
were swept. High SET sensitivity was maintained with the SET
control gate, $V_{\rm C}$, which has negligible coupling to the
dots.

A characteristic feature of a double-dot structure is a
hexagonal-shaped unit cell as the two control gates are swept
\cite{wiel2003}. Each hexagonal cell represents a different charge
state and the dot occupancy is changed by one electron for
adjacent cells. Source(drain) to left(right) dot
($S$($D$)$\leftrightarrow$DOT$_{\rm L(R)}$) transitions are seen
as horizontal(vertical) lines and triple points occur where three
different charge states are degenerate. Lines connecting closest
pairs of triple points denote interdot transitions, completing the
hexagonal cell shape. A charging diagram mapped out from charge
sensing measurements can reveal all of the charge transitions
associated with the double-dot system. Note that this differs from
direct transport measurements where, for example, either the
hexagon cell edges or the triple points can be resolved at
different tunnel barrier conductances and source-drain biases.
Figure 2(a) shows SET conductance as a function of $V_{\rm L}$ and
$V_{\rm R}$ for a device with 600 P donors. In Figure 2(b), a
single trace of SET conductance as a function of $V_{\rm L}$ at
$V_{\rm R}$ = 123 mV shows the characteristic sawtooth behaviour
as the charge transfers are sensed by the SET. In
the charging diagram in Figure 2(a), irregular but repeating cells
are observed. In Figures 3(a) and (b) we zoom in on the corners of
the cells and see that each cell is an elongated hexagon shape and
consistent with a double-dot structure. There is some variability
across devices but several hexagon-shaped cells are observed in
different areas of voltage space across different devices. This
differs from previous work on devices with two isolated dots with
no leads in which parallel charge transfer lines were observed as
electrons were transferred between the two isolated dots
\cite{Buehler}.

We also observe a number of additional charge transfer lines in
the data that are not parallel to each other or part of the
hexagon cells. These are indicative of a more complicated charging
structure and could be due to additional Coulomb blockade events
from random disorder in the device channel, such as charge traps
or unintentional P islands.

The data in Figures 3(a) and (b) show four well-defined charge
states for double-dot devices with 600 and 300 P donors per dot
respectively. In our geometry the SET is nominally centered
between the dots and, as such, we would expect no definition
between the ({\it m}+1, {\it n}) and ({\it m}, {\it n}+1) total
charge states. However, in the 600 donor device, (Figure 3(a)),
some definition between these two states can be observed, due to
the slight misalignment of the SET towards one of the dots.

The elongated dimensions of the hexagon cells are attributed to
the interdot coupling being very much smaller than the coupling
between the dots and gates. The ratio of the mutual capacitance
between the dots, $C_{\rm m}$, to the total dot capacitance,
$C_{\rm L(R)}$, can be extracted by examining the charging diagram
\cite{wiel2003}. Firstly, the direct capacitances of the dots to
their respective gates, $C_{\rm DOT_{L}-V_{L}}$ and $C_{\rm
DOT_{R}-V_{R}}$, were determined from the period of the $S$($D$)
$\leftrightarrow$ DOT$_{L(R)}$ charge state transitions. $C_{\rm
DOT_{L}-V_{L}}$ $\simeq$ 5.8 aF and $C_{\rm DOT_{R}-V_{R}}$
$\simeq$ 5.4 aF. It is important to note that the
cross-capacitances are of significant magnitude, $C_{\rm
DOT_{L}-V_{R}}$ and $C_{\rm DOT_{R}-V_{L}}$ $\sim$ 1-2 aF. This is
due to the reduced size of the device and hence close proximity of
either gate to both dots. The triple-point separations in $V_{\rm
L}$ and $V_{\rm R}$ space, $\triangle V^{\rm m}_{\mathrm{L}}$ and
$\triangle V^{\rm m}_{\mathrm{R}}$ are marked in Figure 3(a). The
ratio of interdot to total dot capacitances are calculated using
the equation $C_{\mathrm{m}}$/$C_{L(R)}$ = $\triangle V^{\rm
m}_{\mathrm{R(L)}}$($C_{\rm DOT_{R(L)}-V_{R(L)}}$/e).
$C_{\mathrm{m}}$/$C_{L}$ $\simeq$ 0.008 and
$C_{\mathrm{m}}$/$C_{R}$ $\simeq$ 0.009, i.e. the interdot
capacitance is $\sim$ 120 times smaller than total dot
capacitances. This imbalance could be addressed if the dots were
closer together or if tunnel barrier control gates were available,
however, at these dimensions, it is problematic to fit so many
individual gates into the device and align them accurately.
Further to this, we note there is a limit on the proximity of the
dots to the leads and to each other in order to maintain well
isolated phosphorus regions. This limit is inherent in the
implantation fabrication process \cite{Jamieson06}.

In both devices, the charge induced on the SET island for the
three different state transitions ($S$ $\leftrightarrow$
DOT$_{L}$, $D$ $\leftrightarrow$ DOT$_{R}$ and DOT$_{L}$
$\leftrightarrow$ DOT$_{R}$) was found to be between 0.02 $e$ and
0.04 $e$, with the variations being due to both misalignment of
the SET and gates to the implanted dots and also to differing
capacitances of the tunnel barriers across the device.

Data from the $N$ = 600 P atom device was modelled using readily
available simulation programs, Crystal-TRIM, FastCap and Simon
\cite{TRIM,Fastcap,Simon}, following similar work in
\cite{Escott}. Crystal-TRIM is a 3D implantation simulator that
predicts the implant profile and peak densities of the nanoscale
dots when given the applied dose, mask geometry and target
material. This package was used as it includes effects, such as
channelling of dopant ions, that result from having a crystalline
implantation target. Using a dose of
4.2~${\times}~10^{19}$~cm$^{-3}$ and 30~nm implant aperture
diameters with a spacing of 100~nm yields an implant profile with
$\sim$ 50~nm dots separated by $\sim$~80~nm in agreement with the
SEM imaging in Figure 1(c). FastCap then uses these dimensions of
the localised, metallic-density phosphorus regions and
characterises the capacitive coupling strength between these
regions and the surface gates. Simon, a Monte-Carlo based single
electron circuit simulator, is subsequently used to model the
effective single electron circuit formed by the capacitively
coupled dots, leads and gates. The gate dimensions used were
measured from SEM images of the device: 50~nm wide gates separated
by 300~nm and a 30~nm wide SET antenna positioned between the
gates. In our devices, transport measurements were precluded by
weak tunnel coupling, but since Simon effectively simulates a
transport measurement, a nominal value for the tunnel junction
resistance that allows transport is required. We
  assumed a value of 1M$\Omega$ and, along with the FastCap capacitances,
  input this into Simon to calculate the stability
  diagram (shown in Figure 4(b)) for a temperature of 100mK. Table 1
compares the modelled capacitance values with the experimental
values extracted from the stability diagram. We found that the
capacitance and cross-capacitance values associated with dot
$\leftrightarrow$ source/drain charge state transitions fell
within the limits of the experimental data error. However, there
is some degree of mismatch between the model and the experimental
data for the ratio of interdot to total dot capacitances,
$C_{\mathrm{m}}$/$C_{L(R)}$. The modelling over estimates
$C_{\mathrm{m}}$/$C_{L(R)}$ by a factor of $\sim$ 10. This could
be due to an overestimation of the size (and hence proximity) of
the implanted dots.

To extend this work further towards devices with potential as
charge or spin qubits, there are a number of strategies to pursue.
The implant dose could be lowered such that fewer P atoms are
implanted into the dots, however, a separate implantation with a
higher dose would be needed to maintain a metallic density of
states in the source and drain leads. At these dimensions,
alignment of the PMMA masks for each implantation would have to be
very precise so as to locate the dot centrally between the leads.
A more elegant solution could be to incorporate a top gate above
the dots and use it to deplete the electrons in the dot. Such a
gate could also be used to control the tunnel barriers between the
dots and the dots and leads, for more precise control and also to
correct for any fabrication misalignments. Another alternative
solution would be to use self-aligning gates made from, for
example, poly-Si that would withstand the high temperature anneal.
These could be designed to firstly form an implantation mask and
then be used as gates which are positioned directly over the dots
and tunnel barriers.

This work has shown that small implanted double-dots in silicon
can be achieved and exhibit classical double-dot
behaviour. Capacitance modelling showed good agreement with the measured data
and device dimensions. There are various ways in which to proceed
towards dots with fewer electrons where observation of excited states
may lead to important coherent measurements, such as the spin states
in silicon. Based on results from GaAs based double-dots, such
measurements could be expected in dots with $N$~$<$ 100 donors.

\ack The authors would like to thank S. E. Andresen, V. Chan and R.
Brenner for helpful discussions and E. Gauja, R. P. Starrett, D.
Barber, G. Tamanyan and R. Szymanski for their technical support.
This work is supported by the Australian Research Council, the
Australian Government and by the US National Security Agency
(NSA), Advanced Research and Development Activity (ARDA) and the
Army Research Office (ARO) under contract number W911NF-04-1-0290.

\newpage
\begin{table}
\begin{center}
  \begin{tabular}{ c c c }
    \hline
     & Experimental & Modelling \\ \hline \hline
    $C_{\rm DOT_{L}-V_{L}}$ (aF) & 5.8 $\pm$ 1.6 & 4.2 \\
    $C_{\rm DOT_{R}-V_{R}}$ (aF) & 5.4 $\pm$ 1.6 & 3.8 \\
    $C_{\rm DOT_{L}-V_{R}}$ (aF) & 1-2 & 1.8 \\
    $C_{\rm DOT_{R}-V_{L}}$ (aF) & 1-2 & 1.7 \\
    $C_{\mathrm{m}}$/$C_{L}$ & 0.008 & 0.09 \\
    $C_{\mathrm{m}}$/$C_{R}$ & 0.009 & 0.13 \\ \hline
  \end{tabular}
\end{center}
\caption{Measured and calculated capacitance and
cross-capacitance values for dot to source/drain transitions and
ratios of interdot to total dot capacitances.}\vspace{1cm}
\end{table}

\begin{figure}[h]
\begin{center}
\includegraphics[width=7.5cm]{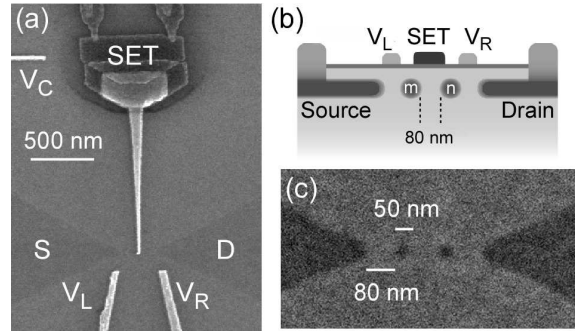}
\caption{(a) SEM image of a Si:P double-dot device with
two surface control gates and SET. (b) Schematic cross-section of
implanted device. (c) SEM image of double-dots and leads after
implantation but before annealing. Regions of dark contrast
indicate implanted phosphorus and pre-anneal damage in the
silicon.}
\end{center}
\end{figure}

\begin{figure}
\begin{center}
\includegraphics[width=5cm]{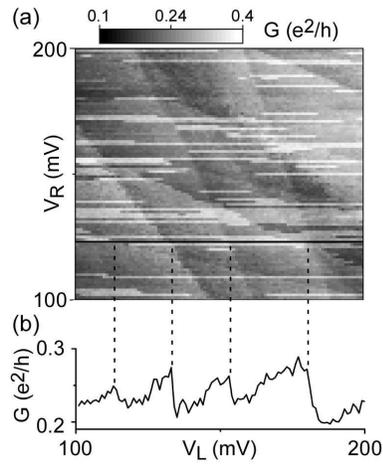}
\caption{(a) Conductance of SET showing charge transfer
events as a function of ${V_{L}}$ and ${V_{R}}$ in an $N$ = 600 P
atom device. Elements of an irregular hexagonal charging diagram
are observed. (b) Single trace of SET conductance plotted as a
function of ${V_{L}}$ at ${V_{R}}$ = 123 mV. A sawtooth signature
is seen as the SET detects each charge transfer.}
\end{center}
\end{figure}

\begin{figure}
\begin{center}
\includegraphics[width=5cm]{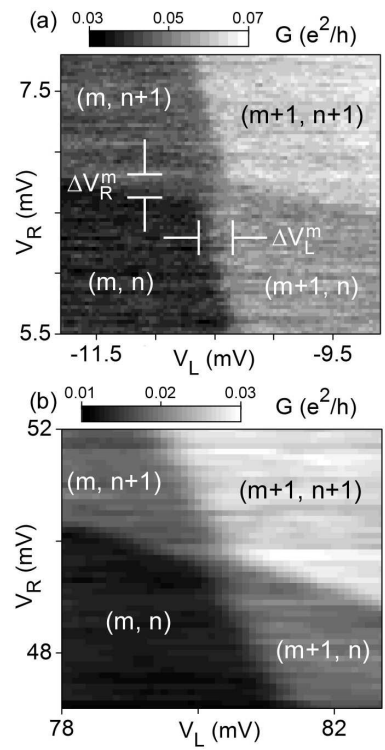}
\caption{(a) and (b) SET conductance for devices with $N$ = 600
and 300 P atoms per dot (respectively) showing hexagonal cells in
the charging diagram. Relative electron population of the dots
within the four charge states are shown.}
\end{center}
\end{figure}

\begin{figure}
\begin{center}
\includegraphics[width=5cm]{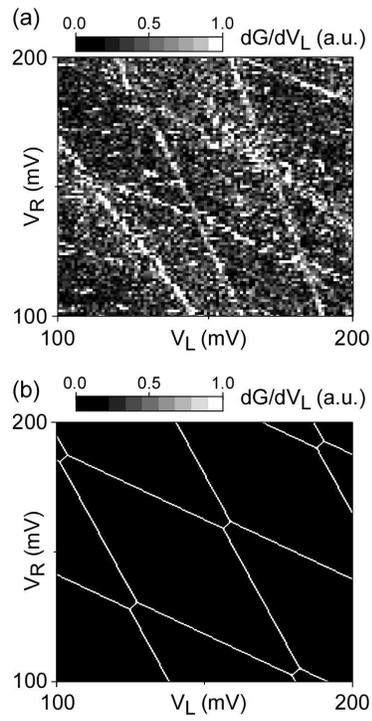}
\caption{(a) Differential conductance of SET as a function of
${V_{L}}$ and ${V_{R}}$ in the $N$ = 600 P atom device.(b)
Numerical modelling of the device in (a).}
\end{center}
\end{figure}

\end{document}